\documentclass[12pt]{iopart}

\usepackage{graphicx}    
\usepackage{multicol}    

\usepackage{amssymb}

\def\3{\ss }           
\def\bsigma{\mbox{\boldmath $\sigma$}}
\DeclareMathSymbol{\blacktriangle}  {\mathord}{AMSa}{"48}  

\begin{document}

\title[Transport coefficients of multi-particle collision algorithms]{\large \bf 
Transport coefficients of multi-particle collision algorithms with velocity-dependent collision rules} 
 
\author{Thomas Ihle} 

\address{Department of Physics, North Dakota State University, P.O. Box 5566,\\
Fargo, ND 58105, USA.}

\address{Max-Planck-Institut f{\"u}r Physik komplexer Systeme, N{\"o}thnitzer Strasse 38,
\\ 01187 Dresden, Germany.} 

\ead{thomas.ihle@ndsu.edu}

\date{\small January 9th, 2008}
 
\vspace{0.1cm}

\begin{abstract} 
Detailed calculations of the transport coefficients of a recently introduced particle-based model for fluid dynamics with 
a non-ideal equation of state are presented. 
Excluded volume interactions are modeled by means of
biased stochastic multiparticle collisions which depend on the local 
velocities and densities. Momentum and energy are exactly conserved locally. 
A general scheme to derive transport coefficients for such biased, velocity dependent collision rules is developed.
Analytic expressions for the self-diffusion coefficient and the shear viscosity are obtained, 
and very good
agreement is found with
numerical results at small and large mean free paths.
The viscosity turns out to be  proportional to the square root of temperature, as in a real gas.
In addition, 
the theoretical framework is applied to a two-component version of the model, and expressions for the viscosity and the difference in diffusion of the two species are given.
\end{abstract}


\section{Introduction}

The interplay between hydrodynamic interactions and thermal fluctuations is crucial for a wide range of phenomena in biophysics and soft-matter physics.
Several particle-based simulation methods such as dissipative particle dynamics \cite{hoog_92,espa_98}, 
smooth particle hydrodynamics \cite{mona_05}, and direct simulation Monte Carlo \cite{bird_76,alex_97} 
have been developed for the efficient modeling of these phenomena
with the motivation to coarse-grain out irrelevant atomistic details while correctly incorporating the essential physics.
One particular method was introduced by Malevanets
and Kapral in 1999 \cite{male_99,male_00a} and is often
called stochastic rotation dynamics (SRD) 
\cite{ihle_01,ihle_03,tuze_03,pool_05,padd_05} 
or multi-particle 
collision dynamics (MPC) \cite{lamu_02,alla_02,ripo_04,wink_04}.
The method is based on so-called fluid particles with continuous positions and 
velocities which follow a simple, artificial dynamics
and undergo efficient multi-particle collisions.
The original SRD algorithm models a fluid with an ideal gas equation of state. 
Recently, the collision rules of the original algorithm have been generalized to model
excluded-volume effects, allowing for a more realistic modeling of dense gases and immiscible binary fluids \cite{ihle_06,tuze_07}.
These new multi-particle collision rules depend on local densities and velocities.
The new models can be thought of as a coarse-grained 
multi-particle collision generalization of a hard sphere fluid since, 
just as for hard spheres, there is no internal energy. 
Their static properties such as the equation of state and the 
entropy density have been derived recently \cite{ihle_06,tuze_07}, and
were shown to be in excellent
agreement with numerical results for the pressure, speed of sound, density fluctuations and the phase diagram.
However,
the dynamic properties are not yet fully understood.
Some preliminary results about the transport coefficients were published previously 
\cite{ihle_06,tuze_07,tuze_06b,ihle_08}, but without derivation
and without results for the viscosity at small mean free path.
In this paper I present the details of how to derive    
expressions for the self-diffusion coefficient and the shear viscosity at small and large mean free path.
For simpler, unbiased collison rules, this was done before \cite{ihle_01,tuze_03,pool_05,ihle_05,ryde_06}; the added difficulty here is that the collision probabilities depend on
the particle velocities which leads to strong correlations.
The derived formulas will be compared
to numerical simulations in various limits. 

The paper is structured as follows: section 2 introduces the simulation model, and
in section 3 the self-diffusion coefficient is derived. In section 4 an equilibrium method to determine
the kinetic part of the viscosity is used: Green-Kubo relations of the kinetic stress correlations are evaluated. 
Section 5 applies a non-equlibrium method to calculate the collisional contribution to the viscosity
by evaluating the momentum transfer in shear flow.
This non-equilibrium method is a more rigorous version of the earlier approach introduced 
in ref.~\cite{ihle_01} and now, in principle, can be used to obtain the shear dependence of the collisional viscosity.
In section 6 the developed scheme is applied to calculate transport coefficients for a binary model and a model with a large collision rate.

\section{Model}

As in the original SRD algorithm, the solvent is modeled by a large number 
$N$ of point-like particles of mass $m_p$ which move in continuous space with 
a continuous distribution of velocities. 
The particle mass is set equal to one throughout this paper.
The system is coarse-grained 
into $(L/a)^d$ cells of a $d$-dimensional cubic lattice of linear 
dimension $L$ and lattice constant $a$. 
The algorithm consists of individual streaming and collision steps. In 
the free-steaming step, the coordinates, ${\bf r}_i(t)$, of the solvent 
particles at time $t$ are updated according to ${\bf r}_i(t+\tau)=
{\bf r}_i(t) + \tau {\bf v}_i(t)$, where ${\bf v}_i(t)$ is the velocity 
of particle $i$ at time $t$ and $\tau$ is the value of the discretized 
time step. In order to define the collision, we introduce a second grid  
with sides of length $2a$ which (in $d=2$) groups four adjacent cells into one 
``supercell''. 
\begin{figure}
\begin{center}
\includegraphics[width=3.0in,angle=0]{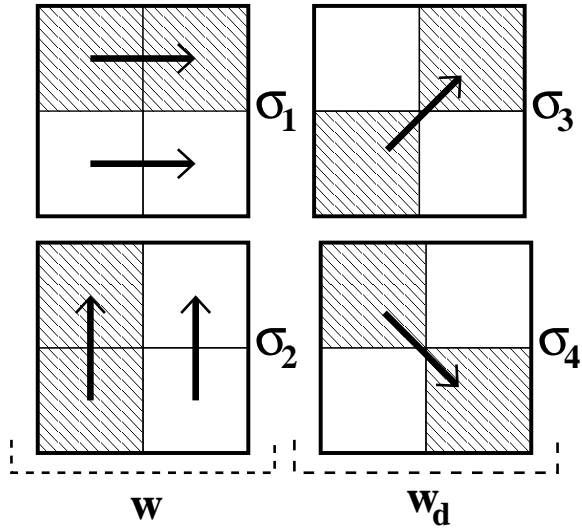}
\caption{
Schematic collision rules. Momentum is exchanged in three ways: a) horizontally along $\bsigma_1$,
b) vertically along $\bsigma_2$, c) diagonally along $\bsigma_3$ and $\bsigma_4$. $w$ and $w_d$ 
denote probabilities of choosing a), b) and c), respectively. From ref. \cite{ihle_06}. 
}
\label{SUPERCOLL}
\end{center}
\end{figure}

As proposed in Refs. \cite{ihle_01,ihle_03}, a random shift of 
the particle coordinates before the collision step is required to ensure 
Galilean invariance. All particles are therefore shifted by the {\it same}
random vector with components in the interval $[-a,a]$ before the collision 
step (Because of the supercell structure, this is a larger interval than in 
the conventional SRD algorithm). 
Particles are then shifted back by the same amount 
after the collision. To initiate a collision, pairs of cells in every supercell 
are randomly selected. As shown in fig. 1, three different choices are 
possible: a) horizontal (with $\bsigma_1=\hat x$), b) vertical ($\bsigma_2
=\hat y$), and c) diagonal collisions (with $\bsigma_3=
(\hat x+\hat y)/\sqrt{2}$ and $\bsigma_4=(\hat x-\hat y)/\sqrt{2}$). Note that 
diagonal collisions are essential to equilibrate the kinetic energies in the 
$x-$ and $y-$directions. 


In every cell, we define the mean particle velocity,
\begin{equation}
{\bf u}_n={1\over M_n}\,\sum_{i=1}^{M_n}\,{\bf v}_i, 
\end{equation}
where the sum runs over all particles, $M_n$, in the cell with index $n$. 
The projection of the difference of the mean velocities of the selected 
cell-pairs on ${\bf \sigma}_j$, 
$\Delta u={\bf \sigma}_j\cdot ({\bf u}_1-{\bf u}_2)$, 
is then used to determine the probability of collision.
If $\Delta u<0$, no collision will be performed. For positive $\Delta u$, a 
collision will occur with an acceptance probability which depends on 
$\Delta u$ and the number of particles in the two cells, $M_1$ and $M_2$.
This rule mimics a hard-sphere collision on a coarse-grained level: For 
$\Delta u>0$ clouds of particles collide and exchange momenta.  
The following acceptance 
probability was introduced in ref. \cite{ihle_06} 
\begin{equation}
\label{NONID0} 
p_A(M_1,M_2,\Delta u)=\theta(\Delta u)\,\,{\rm tanh}(\Lambda) 
 \ \ \ \ {\rm with}\ \ \ \ \Lambda = A\,\Delta u\, M_1M_2 ,   
\end{equation}
where $\theta$ is the unit step function and $A$ is a parameter which 
allows us to tune the equation of state. The hyperbolic tangent function was 
chosen in (\ref{NONID0}) in order to obtain a probability which varies 
smoothly between 0 and 1.
This paper treats the two most relevant limits for the acceptance probability:
In the first one, which was proven to be thermodynamically consistent, one keeps the
collision parameter $A$ small such that the $tanh$ can be replaced by its argument. 
In the other limit, $p_A$ is simply the theta function alone which is equivalent to let 
$A$ go to infinity.
In principle, any other situation with $A<\infty$ can be handled by expanding the $tanh$;
all additional terms still contain solvable integrals. 

Once it is decided to perform a collision, an explicit form for the momentum 
transfer between the two cells is needed. The collision should conserve 
the total momentum and kinetic energy of the cell-pairs participating in 
the collision, and in analogy to the hard-sphere liquid, the collision should 
primarily transfer the 
component of the momentum which is parallel to the connecting vector
$\bsigma_j$. This component will be called the parallel 
or longitudinal momentum. There are many different choices which fulfill these 
conditions. Since the original goal was to obtain a large speed of sound, 
a collision rule was selected which leads to the maximum transfer of the 
parallel component of the momentum and does not change the transverse momentum.
The rule is quite simple; it exchanges the parallel component of the mean 
velocities of the two cells, which is equivalent to a ``reflection'' of 
the relative velocities,  
\begin{equation}
\label{NONID2}
v_i^{\Vert}(t+\tau)-u^{\Vert}=-(v_i^{\Vert}(t)-u^{\Vert})\,, 
\end{equation}
where  $u^{\Vert}$ is the parallel component of the mean velocity of the 
particles of {\it both} cells. The perpendicular component remains 
unchanged, 
\begin{equation}
\label{NONID2A}
v_i^{\perp}(t+\tau)=v_i^{\perp}(t).        
\end{equation}
More specifically, the rule for horizonal collisions in $\bsigma_1$ direction is simply
\begin{equation}
\label{COLLHORI}
\tilde{v}_{i,x}=2u_x-v_{i,x}
\end{equation}
with $\tilde{v}_{i,x}$ denotes $v_{i,x}(t+\tau)$.
The y-components of the particle velocities stay the same, $\tilde{v}_{i,y}=v_{i,y}$.
For the upward diagonal collision along the $\bsigma_3$ direction one has
\begin{eqnarray}
\label{COLLUDIAG}
\nonumber
\tilde{v}_{i,x}=u_x+u_y-v_{i,y} \\
\tilde{v}_{i,y}=u_x+u_y-v_{i,x}
\end{eqnarray}
It is easy to verify that these rules conserve momentum and energy in the 
cell pairs.

Because of $x-y$ symmetry, the probabilities for choosing cell pairs in the 
$x-$ and $y-$ directions (with unit vectors $\bsigma_1$ and $\bsigma_2$, see 
fig. 1) are equal, and will be denoted by $w$.
The probability for choosing diagonal pairs ($\bsigma_3$ and $\bsigma_4$, see  
fig. 1) is given by $w_d=1-2w$. $w$ and $w_d$ must be chosen such that the 
hydrodynamic equations are isotropic and do not depend on the orientation 
of the underlying grid. This was done by considering the temporal 
evolution of the lowest moments of the velocity distribution function, or alternatively, by evaluating the isotropy
of the kinetic part of the viscous stress tensor \cite{tuze_07,ihle_08}.
Both approaches concluded that $w=1/4$ and $w_d=1/2$, and  
previous simulations show that both the speed 
of sound and the shear viscosity are isotropic for this choice. 
Note, however, 
that this does not guarantee 
that all properties of the model are isotropic. This becomes apparent 
at high densities or high collision frequency, $1/\tau\gg 1$, where 
the cubic anisotropy of the grid shows leading to 
inhomogeneous states with cubic symmetry \cite{ihle_06}.
%
%

\section{Diffusion coefficient}

The self-diffusion constant $D$ is given by a sum over the 
velocity-autocorrelation function (eq. (102) in \cite{ihle_03}),
\begin{equation}
\label{DEF_DIFF}
D= \tau\,\left.\sum_{n=0}^\infty\right.' 
\langle v_x(0)\, v_x(n\tau)\rangle\,, 
\end{equation}
where the prime on the sum indicates that the $n=0$ term has the relative weight $1/2$.
The particle mass is set to one throughout the paper.
Assuming molecular chaos, one has
\begin{eqnarray}
\label{CORR_DIFF}
\langle v_x(0)\, v_x(n\tau)\rangle&=&k_B T\,g^n\;\;{\rm where}\;\;\; 
g={\langle v_x(0)\, v_x(\tau)\rangle \over k_B T}
\end{eqnarray}
and the sum (\ref{DEF_DIFF}) turns into a geometric series
which is easily summed up to give
\begin{equation}
\label{GEO_DIFF}
D= {k_BT\,\tau \over 2} 
\,{1+g \over 1-g}\,.
\end{equation}
$g$ is the velocity correlation one time step apart, which completely defines the 
diffusion coefficient in the molecular chaos
approximation.
There is four different contributions to $g$ as a result of possible horizontal, vertical, and the two diagonal 
collisions: 
\begin{equation}
\label{GDET1}
g=w(g_H+g_V)+{w_d\over 2}(g_U+g_D)
\end{equation}

The vertical, $g_V$, and the horizontal contribution, $g_H$, occur with probability $w=1/4$, whereas the upward diagonal
part, $g_U$ and
the downward diagonal part, $g_D$, enter with probability $w_d=1/2$. 
The additional factor $1/2$ in front of the second bracket accounts for the fact that 
the two diagonal collisions are always chosen together and on average only $1/2$ of the particles in a $2a\times 2a$-super cell undergo an upward diagonal collision while the other half is subject to a downward diagonal collision.
In contrast, in a horizontal (or vertical) collision operation {\it all} particles in the super-cell 
can in principle be involved in the collision since the super cell is divided in two double cells with horizontal 
(or vertical) collisions in each one.
The contribution due to vertical collisions is easy to calculate, since in this case, the value of $v_{i,x}$ is unchanged 
in a collision, eq. (\ref{NONID2A}), and $v_{i,x}(\tau)=v_{i,x}(0)$. Following eq. (\ref{CORR_DIFF}) we find 
\begin{equation}
\label{GVERT1}
g_V=
{\langle v_{i,x}(0)\, v_{i,x}(\tau)\rangle \over k_B T}=1
\end{equation}

\subsection{Horizonal collisions}

Consider now the horizontal contribution, $g_H$. We assume that there are $n$ particles in the left subcell, and $m$ particles
in the right subcell of the cell pair where the collision happens.
It is also assumed, that the tagged particle $i$ is contained in the left subcell.
For the moment the particle numbers $n$ and $m$ are kept fixed.
According to eq. (\ref{NONID0}), for small $A$, a collision occurs with the acceptance probability, 
\begin{equation}
\label{PACCEPT}
p_A=A\,n\,m\;\theta(\Delta u)\;\Delta u\,;
\end{equation}
and the value of $v_{i,x}$ changes in the following way 
$v_{i,x}(\tau)=2u_x-
v_{i,x}(0)$.
Here, $u_x$ is the x-component of the mean velocity of {\it both} subcells:
\begin{equation}
\label{UXDEF}
u_x={1\over n+m} \sum_{j=1}^{n+m}\,v_{j,x}
\end{equation}
For horizontal directions, $\Delta u$ contains only the x-components of the particle velocities:
\begin{equation}
\label{DUDEF}
\Delta u={1\over n} \sum_{j=1}^n\,v_{j,x}-
{1\over m} \sum_{j=n+1}^{n+m}\,v_{j,x}\,.
\end{equation}.

If a collision is not accepted, 
which happens with probability $1-p_A$, the value of $v_{i,x}$ remains unchanged: $v_{i,x}(\tau)=v_{i,x}(0)$.
From these considerations it follows,
\begin{equation}
\label{GHCALC1}
g_H={1 \over k_B T} \langle p_A\,v_{i,x} (2u_x-v_{i,x})+(1-p_A)v_{i,x}^2\rangle\,.
\end{equation}
The average is performed over the velocity distribution (at time $t=0$) of all involved particles and over the particle 
numbers in the two subcells forming a collision cell.
Note that there is a non-trivial coupling between the velocity-dependent acceptance probability $p_A$ and the updated velocity
$v_{j,x}(\tau)$. Ignoring this coupling, which amounts to the approximation $\langle p_A\, v_{j,x}(0) v_{j,x}(\tau)
 \rangle\approx \langle p_A \rangle\langle v_{j,x}(0) v_{j,x}(\tau) \rangle$
leads to an errorous result which is a factor of about two smaller than the correct one, even for large particle numbers $n$, $m$.

Using $\langle v_{i,x}^2\rangle=k_B T$, eq. (\ref{GHCALC1}) simplifies to
\begin{equation}
\label{GHCALC2}
g_H=1+{2 \over k_B T} \langle p_A\,v_{i,x} (u_x-v_{i,x})\rangle\,
=1+{2 \over k_B T} \langle K_{nm}\rangle_{n,m}
\end{equation}
The average $\langle ... \rangle_{n,m}$ denotes an average over the particle number distribution in the two subcells (which are Poisson-distributed and uncorreletad in an ideal gas).
$K_{nm}$ contains the average of the velocity distribution, which is Maxwellian if molecular chaos is assumed; and it has to be split up in two parts,
\begin{equation}
\label{GHCALC3}
K_{nm}={1 \over 2}\left( K_{nm}^l+K_{nm}^r\right)
\end{equation}
because there are two possibilities occuring with probability $1/2$, respectively : (i) the tagged particle $i$ can be in the left subcell
as one of the $n$ particles there, or (ii) can be in the right subcell with a total of $m$ particles.
Hence, one has
\begin{eqnarray}
\nonumber
K_{nm}^l&=&A\,n\,m\,J_{nm}^l\;\;\;{\rm with} \\
\label{GHCALC4}
J_{nm}^l&=&\int_{-\infty}^{+\infty}\,dv_{i,x}^{n+m}\,\theta\left( \Delta u \right)\,\Delta u\,
v_{1,x}(u_x-v_{1,x})\,\prod_{i=1}^{n+m} \,f_0(v_{i,x})\,.
\end{eqnarray}
$f_0(v_{i,x})$ is the Maxwell-Boltzmann distribution,
\begin{equation}
\label{GHCALC5}
f_0(v_{i,x})=\Gamma\,{\rm exp}\left(-{v_{i,x}^2 \over 2 k_B T}\right)\,,\;\;\;{\rm with}\;\;\Gamma={1\over \sqrt{2\pi k_B T}}\,.
\end{equation}
Assuming the tagged particle to be in the left subcell in the definition of $J_{nm}^l$ means also that $n>0$
in the expression for $J_{nm}^l$. For the same reason, $m$ must be larger than zero in expressions for $J_{nm}^r$.

Because of the $\theta$-function the direct evaluation of the $(n+m)$-dimensional integral becomes very tedious for large $n,m$.
Therefore, we express the $\theta$-function by means of the $\delta$-function,
\begin{equation}
\label{GHCALC6}
\theta(\Delta u)=\int_0^{\infty}\,\delta(\Delta u-c)\,dc
\end{equation}
which allows us to use the integral-representation of the $\delta$-function, 
\begin{equation}
\label{GHCALC7}
\delta(\Delta u-c)=\int_{-\infty}^{+\infty}\,{\rm exp}(ik(\Delta u-c))\,{dk\over 2 \pi}
\end{equation}
Even though this transformation leads to two more integrations it enables a straightforward calculation of $J_{nm}^l$
for arbitrary $n$ and $m$.
It involves only simple integrals of type $\int _0^{\infty} x^n\,exp(-x^2)$ with $n=0,1,2,3$. 
The integral to be solved is:
\begin{eqnarray}
\nonumber
& &J_{nm}^l=\Gamma^{n+m} \int_0^{\infty}\,dc\,\int_{-\infty}^{\infty}\,{dk\over 2 \pi}   
\int_{-\infty}^{+\infty}\,dv_{i,x}^{n+m} \\
\nonumber
&\times& {\rm exp}\left[ik\left(  
{1\over n} \sum_{j=1}^n\,v_{j,x}-
{1\over m} \sum_{j=n+1}^{n+m}\,v_{j,x}-c\right) 
-{1\over 2 k_B T} (v_{1,x}^2+...+v_{n+m,x}^2)\right] \\
\label{GHCALC8}
&\times & 
v_{1,x}\,\left({1\over n} \sum_{j=1}^n\,v_{j,x}-{1\over m} \sum_{j=n+1}^{n+m}v_{j,x}\right)
\,\left(
{1\over n+m} \sum_{j=1}^{n+m}\,v_{j,x}
 -v_{1,x}\right)\,.
\end{eqnarray}
First, the argument of the exponential has to be transformed into a sum of squares in order to decouple the velocities and to 
enable integrating over them.
This is achieved by the transformations for the velocities in the left subcell:
\begin{equation}
w_i=v_{i,x}-{k_B T ik\over n}\;\;{\rm for}\;i=1...n 
\end{equation}
and for the ones in the right subcell,
\begin{equation}
w_i=v_{i,x}+{k_B T ik\over m}\;\;{\rm for}\;i=n+1...n+m\,.
\end{equation}
Expressing eq. (\ref{GHCALC8}) in the new variables $w_i$ gives:
\begin{eqnarray}
\nonumber
J_{nm}^l&=&\Gamma^{n+m} \int_0^{\infty}\,dc\,\int_{-\infty}^{\infty}\,{dk\over 2 \pi}   
\int_{-\infty}^{+\infty}\,dw_i^{n+m} \\
\nonumber
&\times& {\rm exp}\left(-ikc-{k_B T \gamma k^2\over 2}\right)
\,{\rm exp}\left({1\over 2 k_B T} \left[ w_1^2+w_2^2+...+w_{n+m}^2\right] \right) \\
\nonumber
& &\left( {w_1+w_2+...+w_n\over n} -{w_{n+1}+w_{n+2}+...+w_{n+m}\over m} + k_B T ik\gamma \right) \\
& &\left(w_1+{k_B T ik\over n}\right)
\left({w_1+w_2+...+w_{n+m}\over n+m} -w_1-{k_B T ik\over n} \right)
\label{GHCALC9}
\end{eqnarray}
with $\gamma=(1/n)+(1/m)$.
The integration over the $w_i$ is straightforward and yields
\begin{eqnarray}
\nonumber
J_{nm}^l&=&\Gamma^{n+m} \int_0^{\infty}\,dc\,\int_{-\infty}^{\infty}\,{dk\over 2 \pi}   
\\
\nonumber
\label{GHCALC10}
& & \times {\rm exp}\left(-ikc-{k_B T \gamma k^2\over 2}\right) 
\left[ {(k_B T)^3ik^3\gamma\over n^2}+S (k_B T)^2ik \right] \\ 
S&=&-{2\over n^2}+\gamma\left({1\over n+m}-1\right)
\label{GHCALC11}
\end{eqnarray}
The remaining integral over $h$ and $k$ is done easily using similiar transformations. The final result is:
\begin{equation}
\nonumber
J_{nm}^l=-{(k_B T)^{3/2}\over \sqrt{2\pi\gamma}}\left[{1\over n^2}+\left( {1\over n}+{1\over m}\right)\left( 1-{1\over n+m}
\right)\right]
\label{GHCALC12}
\end{equation}
which is always negative.

The quantity $J_{nm}^r$ is given by an expression almost identical to eq. (\ref{GHCALC12}), 
one merely has to replace $v_{1,x}$ by $v_{n+1,x}$ at the two positions where $v_{1,x}$ appears isolated. This changes the index of the tagged particle from $1$ to $n+1$, meaning that
this particle is not in the left subcell but in the right one.
One finds $J_{nm}^r=J_{mn}^l$ leading to an expression for $J_{nm}$ which is symmetric in $n$ and $m$:
\begin{equation}
\nonumber
K_{nm}={Anm\over 2}(J_{nm}^l+J_{nm}^r)=
-{A(k_B T)^{3/2}\over \sqrt{2\pi\gamma}}\left[
{n^2+m^2\over 2 nm}+n+m-1
\right]
\label{GHCALC13}
\end{equation}
Using eqs. (\ref{GHCALC2}-\ref{GHCALC4})
$g_H$ is calculated.
In an ideal gas, $K_{nm}$ has to be averaged over the poisson-distributed particle numbers in the subcells, $n$ and $m$.
In the current model with a non-ideal equation of state, particle number fluctuations are supressed compared to an ideal gas.
Therefore, we will neglect these fluctuations completely for the moment, and replace $n$ and $m$ by the mean number of particles in a subcell $M=\langle n\rangle=\langle m\rangle$. 
This leads to the final result
\begin{equation}
\label{GHFINAL}
g_H=1-2 A \sqrt{{k_B T\over \pi}}
M^{3/2}
\end{equation}

\subsection{Diagonal collisions}

Here, the upward diagonal collision (along $\bsigma_3$ in fig. \ref{SUPERCOLL}) is considered first.
In this case, the velocity difference, $\Delta u$, depends on both the x- and the y-component of the particle velocities,
\begin{equation}
\label{DELTAUDIAG}
\Delta u=\bsigma_3\cdot \left({\bf u}_1-{\bf u}_2\right)={1\over \sqrt{2}}
\left(u_{1,x}-u_{2,x}+u_{1,y}-u_{2,y}\right)
\end{equation}
with the normal vector of the diagonal collision direction,
$\bsigma_3=(1,1)/\sqrt{2}$,
leading to a more involved calculation of $g_U$.
Furthermore, both components of the velocities change in a collision, but only the change of the x-component is needed:
\begin{equation}
\nonumber
v_{i,x}(\tau)=u_x+u_y-v_{i,y} 
\end{equation}
This change occurs with probability $p_A$, see eq. (\ref{PACCEPT}), but with $\Delta u$ given in 
eq. (\ref{DELTAUDIAG}).
It follows
\begin{equation}
\label{GDCALC1}
g_U={1 \over k_B T} \langle p_A\,v_{i,x} (u_x+u_y-v_{i,y})+(1-p_A)v_{i,x}^2\rangle\,.
\end{equation}
The average must be done over the velocity distribution (at time $t=0$) of all involved particles and over the particle number in the two subcells forming a collision or double-cell.
Because of $\langle v_{i,x}^2\rangle=k_B T$ we can write
\begin{equation}
\label{GDCALC2}
g_U=1+{1\over k_B T}\langle R_{nm}-L_{nm}\rangle_{n,m}
\end{equation}
with the abbrevations
\begin{eqnarray}
R_{nm}&=&\langle p_A\,v_{i,x} (u_x+u_y-v_{i,y})\rangle_T \\
L_{nm}&=&\langle p_A\,v^2_{i,x}\rangle_T
\end{eqnarray}
The subscript $T$ means that only the thermal average is taken; the particle numbers $n$ and $m$ of the subcells are 
kept fixed in this average.

We consider again the two equivalent situations that the tagged particle is in the left subcell with a total
of $n$ cells denoted by the superindex $l$, or in the right cell, described by the superindex $r$.
Both possibilities occur with equal probability $1/2$ and we have
\begin{eqnarray}
\nonumber
R_{nm}&=&{1\over 2}\left(R_{nm}^l+R_{nm}^r \right) \\
L_{nm}&=&{1\over 2}\left(L_{nm}^l+L_{nm}^r \right)
\end{eqnarray}
Other abbreviations are introduced as $R_{nm}=A n m\, F_{nm}$, and
$L_{nm}=A n m\, I_{nm}$.
This time the y-component of the velocities shows up in the calculations, leading to a 
$2n+2m$-dimensional integral for $F_{nm}^l$.
We again adopt the trick to express the $\theta$-function in the acceptance probability $p_A$ by means of an 
integral over the $\delta$-function
and use the integral representation of $\delta(x)$, which yields the following integral:
\begin{eqnarray}
\nonumber
F_{nm}^l&=&\Gamma^{n+m} \int_0^{\infty}\,dc\,\int_{-\infty}^{\infty}\,{dk\over 2 \pi}   
\int_{-\infty}^{+\infty}\,dv_{i,x}^{n+m}\, 
\int_{-\infty}^{+\infty}\,dv_{i,y}^{n+m} \\
\nonumber
&\times& {\rm exp}\left\{ik\left({1\over \sqrt{2}}\left[  
{1\over n} \sum_{j=1}^n\,(v_{j,x}+v_{j,y})-
{1\over m} \sum_{j=n+1}^{n+m}\,(v_{j,x}+v_{j,y})
\right]-
h\right)\right\} \\ \nonumber 
&\times& {\rm exp}\left\{-{1\over 2 k_B T} \left(v_{1,x}^2+...+v_{n+m,x}^2+
v_{1,y}^2+...+v_{n+m,y}^2\right)\right\} \\
\nonumber      
&\times & 
{v_{1,x}\over \sqrt{2}} \,
\left({1\over n} \sum_{j=1}^n\,(v_{j,x}+v_{j,y})-{1\over m} \sum_{j=n+1}^{n+m}(v_{j,x}+v_{j,y})
\right)
\\
\label{GDCALC8}
&\times& \left(
{1\over n+m} \sum_{j=1}^{n+m}\,(v_{j,x}+v_{j,y})
 -v_{1,y}\right)\,.
\end{eqnarray}
This integral is solved straightforwardly in a very similar fashion to the one in eq. (\ref{GHCALC8}).
The final result is
\begin{equation}
\label{GDCALC9}
F_{nm}^l={(k_B T)^{3/2}\over \sqrt{2\pi \gamma}}\left[{1\over nm}-{1\over 2 n^2}\right]
\end{equation}
A similar calculation leads to
\begin{equation}
I_{nm}^l={(k_B T)^{3/2}\over \sqrt{2\pi \gamma}}\left[\gamma+{1\over 2 n^2}\right]
\end{equation}
Again, we observe the symmetry that the quantities with upper index $r$ follow from the ones with index $l$
by simply interchanging $n$ and $m$ in the obtained expressions: $F_{nm}^r=F_{mn}^l$, $I_{nm}^r=I_{mn}^l$.
Inserting these results into eq. (\ref{GDCALC2}) and neglecting particle number fluctuations gives
\begin{equation}
\label{GDFINAL}
g_U=1-A \sqrt{{k_B T\over \pi}}M^{3/2}
\end{equation}

The evaluation of the downward diagonal contribution was done in a similar way, and for symmetry reasons 
one finds
$g_D=g_U$.
According to eqs. (\ref{GHFINAL}), (\ref{GDFINAL}) and (\ref{GEO_DIFF}), (\ref{GDET1}) 
the diffusion coefficient follows in the limit of small collision parameter $A$ as
\begin{equation}
\label{EXACT_DIFF}
D=k_B T\,\tau\left( {1\over A}\,\sqrt{\pi \over k_B T}\;M^{-3/2}  
-{1 \over 2} \right)\,,
\end{equation}
which is in good agreement with simulation data, see fig. 2.
It is interesting to note, that the contributions to $D$ from the vertical and horizontal collisions together,
$(g_H+g_V)/4$, are exactly equal to the ones from the diagonal ones, $(g_U+g_D)/4$, which 
is a sign of the correct isotropic choice of the probabilities $w$ and $w_d$.

\subsection{Density fluctuations}
Density fluctuations were completely ignored in the derivation of eq. (\ref{EXACT_DIFF}).
In principle, the correct density fluctuations which follow from the equation of state could be incorporated,
but here we will only estimate 
the effect of these fluctuations.
This can be done by assuming that the particle numbers $n$ and $m$ in the two subcells are uncorrelated and behave like in an ideal gas.
Instead of replacing $n$ and $m$ by the mean value $M$ in the expressions for $g_H$ and $g_U$ we average over a 
Poisson-distribution.
In this case the probability to find $n$ particles in the left subcell is equal to 
\begin{equation}
p_{cell,n}={\rm exp}(-M)\,M^n/n!\,,
\end{equation}
where
$M$ is the average particle number per cell, $M=\langle n\rangle$.
On the other hand, any one of these $n$ particles can be our tagged particle of index $i$.
Hence, the probability that a given particle is in a subcell with a total of $n$ particles is equal to
$p_n\sim n\,p_{cell,n}$ (see for instance Ref. \cite{pool_05}).
Normalizing this probability gives
\begin{equation}
\label{PROBDEF}
p_n={1\over M}{{\rm e}^{-M}\,M^n\over (n-1)!}
\end{equation}
Care must be taken for the case when at least one of $n$ or $m$ are zero.
The formal expression might look divergent, but one has to keep in mind how the algorithm is actually working:
The acceptance probability $p_A\sim n m$ is zero in this case, hence no collision is executed.
This means that the velocities do not change, $v_{i,x}(\tau)=v_{i,x}(0)$, and the actual value of $g_H$ and $g_U$ is one.
This can be incorporated by starting the summations over $n$ and $m$ at $n=1$, $m=1$ instead of zero in eq. 
(\ref{GH_FLUC}).

There is another subtle point about this average:
For quantities with index $l$, such as $K_{nm}^l$, 
we have to use $p_n$ to average over the particle number in the left subcell which contains the tagged particle,
but $p_{cell}$ to average over the particle number $m$ in the right subcell.
This is because no specific particle is adressed in the right subcell; all what is needed
is the probability to find a given particle number $m$ in this cell.
These arguments lead to the following result, 
\begin{eqnarray}
\nonumber
g_H&=& 1-2A\sqrt{k_B T\over 2\pi}\left< \sqrt{nm\over n+m}\left[ {m\over 2n}+{n\over 2m}+n+m-1\right] \right>_{n,m} \\
\nonumber
&=& 1-2A\sqrt{k_B T\over 2\pi} \sum_{n=1}^\infty\,\sum_{m=1}^\infty \left\{ 
p_n p_{cell,m} \sqrt{nm\over n+m} \left[ {m\over 2n}+{1\over 2}(n+m-1)\right] 
\right.
\\
\label{GH_FLUC}
& &+\left. p_m p_{cell,n}  \sqrt{nm\over n+m} \left[ {n\over 2m}+{1\over 2}(n+m-1)\right] 
\right\} 
\end{eqnarray}
Similar expressions hold for the diagonal terms, $g_U$ and $g_D$.
These averages over the Poisson-distributed particle numbers cannot be done analytically.
Therefore, a numerical evaluation of the sums was
performed and inserted in expressions (\ref{GEO_DIFF},\ref{GDET1}) to obtain the diffusion coefficient.
This result provides an upper limit for the effect of fluctuations. A check shows that it only differs by a few 
percent from the result without fluctuations.
As seen in figure 2 of ref. \cite{ihle_06}, the simulation results 
show excellent agreement
with the formula without particle number fluctuations.
The effective excluded volume interactions of the current non-ideal model lead to a supression of density fluctuations.
Hence, it seems plausible that neglecting particle number fluctuations entirely is a better approximation for this  model
than assuming strong ideal-gas like density fluctuations.




\section{Kinetic viscosity}

The multi-particle collision algorithm consists of two steps -- streaming and collision.
Each of these steps redistribute momentum. 
At large mean free path $\lambda$ compared to the cell size $a$, this transport mainly occurs in the streaming step 
where momentum is directly carried away by the moving particles.
At small mean free path, most momentum is transfered in the collision step.
The momentum transport during streaming is characterized by the so-called kinetic viscosity, whereas the 
transfer during collisions gives rise to the collisional contribution to the viscosity. 
The kinetic viscosity is given by the Green-Kubo relation, 
\cite{ihle_05}, 
%
%
\begin{equation}
\label{VISC1}
\nu_{kin}={\tau\over N k_B T}{\sum_{n=0}^{\infty}}'\langle\sigma_{xy,k}(0)\sigma_{xy,k}(n\tau)\rangle\,,
\end{equation}
The prime on the sum indicates that the first $n=0$ term has a prefactor $1/2$. $\sigma_{xy,k}$ is the off-diagonal element of the kinetic stress tensor,
\begin{equation}
\label{VISC2}
\sigma_{xy,k}=\sum_{j=1}^N\,v_{j,x} v_{j,y}\,.
\end{equation}
In order to evaluate eq. (\ref{VISC2}) the correlation of the stress tensor for only one time step apart is needed,
since under the assumption of molecular chaos, the sum in eq. (\ref{VISC1}) is again a geometric series. 
Hence, the following quantity is required,
\begin{equation}
b=\sum_{i,j=1}^N\langle v_{i,x}(0)v_{i,y}(0)v_{j,x}(\tau)v_{j,y}(\tau)\rangle\,.
\end{equation}
Analogous to eq. (\ref{GDET1}) we have
\begin{equation}
b=w(b_V+b_H)+{w_D\over 2}(b_U+b_D)
\end{equation}
where $b_V$ describes the contribution from the vertical collisions, $b_H$ the horizontal, and $b_U$ and $b_D$
the ones from the diagonal collisions.
\vspace{0.4cm}

\subsection{Horizontal contribution}

During a horizontal collision, only $v_{j,x}$ can change. Thus, using 
$\langle v_{i\alpha}v_{j\beta}\rangle=\delta_{ij}\delta_{\alpha\beta}\,k_B T$ we have
\begin{eqnarray}
b_H&=& \sum_{i,j=1}^N\langle \left[ p_A v_{i,x}v_{i,y}[2u_x-v_{j,x}]v_{j,y}
+(1-p_A)v_{i,x}v_{i,y}v_{j,x}v_{j,y}\right]\rangle \\
&=&N(k_B T)^2+2k_B T\left\langle p_A \sum_{j=1}^N\,v_{j,x}(u_x-v_{j,x})\right\rangle
\end{eqnarray}
where the time argument was omitted since all velocities are now at the same time $t=0$.
The sum over all particles can be rewritten as a sum over all double cells and over the particles in each of them
since both $u_x$ and $p_A$ depend only on the properties of the particles in the considered double cell.
This leads to the following simplification:
\begin{equation}
b_H=N(k_B T)^2+2k_B T N_H \left\langle p_A \sum_{j=1}^{n+m}\,v_{j,x}(u_x-v_{j,x})\right\rangle
\end{equation}
$N_H=N/(2M)$ is the number of double cells for horizontal collisions. 
The sum runs over the $n$ particles in the left subcell
of a double cell and over the $m$ particles in the right subcell.
The average has to be taken over $n$, $m$ and over the velocity distribution.
Inserting the definition of $p_A$, eq. (\ref{PACCEPT}), yields
\begin{equation}
\label{VISC_K}
b_H=N(k_B T)\left\{k_B T+{A\over M}\left\langle nm \sum_{j=1}^{n+m}\,\langle \theta(\Delta u) \Delta u\,
v_{j,x}(u_x-v_{j,x})
\rangle_T\right\rangle_{n,m}
\right\}
\end{equation}
As before, the subindex $T$ denotes the thermal average at fixed particle numbers per cell, and the subindex
$n,m$ refers to the average over the particle numbers.

The remarkable feature of eq. (\ref{VISC_K}) is that the calculation of $b_H$ is reduced to the same
integrals which appeared 
in the previous section in the calculation of the diffusion coefficient, namely the quantities $K_{nm}^l$
and $K_{nm}^r$ defined in eqs. (\ref{GHCALC2},\ref{GHCALC3}).
This also means as in the case of the diffusion coefficient, there again is a non-trivial 
coupling between the velocity-dependent acceptance probability and the elements of the stress tensor.
As before, a mean-field like treatment consisting of multiplying the averaged value of $p_A$ with the 
averaged stress-correlations leads to a large error -- a factor of two. This error persists even in the case of high particle numbers
where one naivly would assume this decoupling to be correct.
According to eq. (\ref{GHCALC3},\ref{VISC_K}) we have
\begin{equation}
\label{VISC_K1}
b_H=N(k_B T)\left\{k_B T+{A\over M}\langle nm 
\left( n K_{nm}^l+m K_{nm}^r\right)
\rangle_{n,m}
\right\}
\end{equation}
Using the explicit form of the $K's$, eqs. (\ref{GHCALC3},\ref{GHCALC4}), gives 
\begin{equation}
\label{VISC_K2}
b_H=N(k_B T)^2\left\{1-{A\over M}\sqrt{k_B T\over 2\pi} \langle  
(n+m)\sqrt{nm(n+m)}
\rangle_{n,m}
\right\}
\end{equation}
Ignoring particle number fluctuations, and replacing $n$ and $m$ by their mean value $M$ leads to the final
expression:
\begin{equation}
\label{VISC_K3}
b_H=N(k_B T)^2\left\{1-2A  \sqrt{k_B T\over \pi} M^{3/2}\,,  
\right\}
\end{equation}
which looks identical to $g_H$, eq. (\ref{GHFINAL}), except the prefactor.

Since the stress tensor is symmetric with respect to interchanging $x$ and $y$, and the collision rules are 
constructed to be also x-y-symmetric, one has
for the vertical contribution: $b_V=b_H$.

\subsection{Diagonal contributions}

The upward diagonal contribution can be written as
\begin{eqnarray}
b_U&=& \sum_{i,j=1}^N\left\langle\left[ p_A v_{i,x}v_{i,y}\tilde{v}_{j,x}\tilde{v}_{j,y}
+(1-p_A)v_{i,x}v_{i,y}v_{j,x}v_{j,y}\right]\right\rangle \\
&=&N(k_B T)^2+
\sum_{i,j=1}^N\left\langle p_A v_{i,x}v_{i,y}[\tilde{v}_{j,x}\tilde{v}_{j,y}-v_{j,x}v_{j,y}]
\right\rangle
\end{eqnarray}
The tilde denotes velocities at time $\tau$ as opposed to time zero.
The sum over the particle index $j$ is rewritten as a sum over all available double cells
for diagonal collisions and another sum over the particles in that cell.
The reason again is that both the velocities at time $\tau$ and the collision probability $p_A$ depend only
on the considered double cell. One finds,
\begin{equation}
b_D=N(k_B T)^2+N_D
\left\langle \left( \sum_{i=1}^N\, v_{i,x}v_{i,y}\right) 
 \sum_{j=1}^{n+m} p_A [\tilde{v}_{j,x}\tilde{v}_{j,y}-v_{j,x}v_{j,y}]
\right\rangle
\end{equation}
where $N_D=N/(4M)$ is the number of double cells for diagonal collisions.
(This number is smaller than the one for horizontal collisions due to the specific algorithm chosen)
Inserting the collision rules for upward diagonal collisions, 
eq. {\ref{COLLUDIAG}), yields
\begin{eqnarray}
\nonumber
& &b_D=N(k_B T)^2+ \\
& &N_D\left\langle \left( \sum_{i=1}^N\, v_{i,x}v_{i,y}\right) 
\sum_{j=1}^{n+m} p_A (u_x+u_y)[u_x+u_y-v_{j,x}-v_{j,y}]
\right\rangle \\
&=&N(k_B T)^2+
{A N\over 4 M} \left\langle \left( \sum_{i=1}^N\, v_{i,x}v_{i,y}\right) 
\left\langle nm\sum_{j=1}^{n+m} \theta(\Delta u)\, \Delta u\right.\right.\\
& &\left.\left.\times (u_x+u_y)[u_x+u_y-v_{j,x}-v_{j,y}]
\right\rangle_T\right\rangle_{n,m}
\end{eqnarray}
$u_{\alpha}$ and $\Delta u$ are properties of the collision cell and can be put in front of the sum.
The summation gives zero, since $\sum^{n+m}\,(u_\alpha-v_{j,\alpha})=0$.
Thus,
\begin{equation}
\label{BURES}
b_U=N (k_B T)^2
\end{equation}
For the downward diagonal collisions a similar calculation results in
$b_D=b_U$ for symmetry reasons.

\vspace{0.4cm}

{\bf Final result for kinetic viscosity}

Combining the results for $b_V=b_H$, eq. (\ref{VISC_K3}), and $b_D=b_U$, eq.(\ref{BURES}), gives
\begin{equation}
b={1\over 4}(2 b_H)+{1\over 4} (2 b_D)=N(k_B T)^2\left(1-A\sqrt{k_B T\over \pi} M^{3/2}\right)
\end{equation}
With $r=b/(N (k_B T)^2)$
the summation of the geometric series leads to the kinetic viscosity
\begin{equation}
\label{VISCOS_KIN_A}
\nu_{kin}={\tau k_B T\over 2} {1+r\over 1-r}=\tau k_B T
\left({1\over A}\sqrt{\pi\over k_B T} M^{-3/2}-{1\over 2}\right)
\end{equation}
which is exactly equal to the diffusion coefficient, eq. (\ref{EXACT_DIFF}) and also scales with $\sqrt{k_B T}$.
Hence, in the limit of large mean free path
the Schmidt number, $Sc=\nu/D$, of this model goes to one.
$Sc$ can be modified by adding additional regular SRD rotations on the subcell level which do not change
the equation of state.

\section{Collisional viscosity}

In contrast to the kinetic viscosity, the collisional contribution is evaluated in non-equilibrium
--- in shear flow with shear rate $\dot{\gamma}$.
Only the limit of infinitesimally small shear rates will be considered, that means the collisional viscosity will be calculated to zeroth order in $\dot{\gamma}$.
In principle, it is possible to perform the calculation in higher order which would tell us about eventual shear- or shear thinning similar to ref. \cite{ryde_06,ryde_05}, however, in lowest order a number of convenient 
approximations can be made.
For example, we can neglect the deviation of the particle velocity distributions from a Gaussian shape
since this distortion is caused by the very small shear rate and gives rise to a higher order term 
in the viscosity.
Due to the shear, the velocity distribution depends on the height $y_i$ of particle $i$ and the following factorized form
of the N-particle distribution function can be assumed,
\begin{eqnarray}
\label{DEFFNX}
f_{N,x}&\approx&\prod_{i=1}^N f_0(v_{i,x}-\dot{\gamma}y_i) \\
\label{DEFFNY}
f_{N,y}&\approx&\prod_{i=1}^N f_0(v_{i,y}) 
\end{eqnarray}
where $f_0$ is the Maxwell-Boltzmann distribution, eq. (\ref{GHCALC5}).
This factorization is equivalent to assuming {\em Molecular chaos} 
which was shown to be a very good approximation for calculations of the collisional
part of transport coefficients \cite{ihle_05}.
A given collision cell (which consists of two subcells) is divided by a 
line $y=h$, and average transfer of x-momentum  across this line is calculated.
Vertical collisions do not change the x-component of the velocities at all, and hence do not have to be considered. However, both diagonal and the horizontal collisions transfer x-momentum across this line and will
be evaluated in the next subsections.

\subsection{Horizontal collisions}

A dividing line is assumed to go across a pair of horizontally aligned subcells at fixed height $y=h$
with $0\le h \le a$, see fig. \ref{BOXPART}.
The left subcell contains $N_A$ particles which are split into two populations below and above the dividing line
with particle numbers $n$, $m$, respectively, such that $N_A=n+m$.
The line divides the particle population in the right subcell into $p$ and $q$ particles, with total particle number
$N_B=p+q$.
For now, the particle numbers $n$, $m$, $p$, and $q$ are kept fixed.
I consider the change of total x-momentum of the $n+p$ particles {\it below} the dividing line, 
which equals the amount of x-momentum transferred over this line to the particles denoted as filled circles
in fig. \ref{BOXPART}.
This momentum transfer $\Delta p_x$ for fixed height $h$ and fixed particle numbers is averaged over the particle positions 
and velocities resulting in
\begin{equation}
\label{COLLHOR1}
\langle{\Delta} p_x\rangle_{nmpq}^H=\left\langle \left\langle  
p_A\,\sum_{j=1}^{n+p} \Delta v_{i,x} \right\rangle_V \right\rangle_Y\,.
\end{equation}
The upper index, $H$, stands for ``horizontal'' and the averages are defined as:
\begin{eqnarray}
\label{AVERV}
\langle ... \rangle_V&=&\int_{-\infty}^{\infty}\,...f_{N,x}\,dv_{i,x}^{n+m+p+q} \\ 
\label{AVERY}
\langle ... \rangle_Y&=&{1\over h^{n+p}}\int_0^h\,dy_i^{n+p} {1\over (a-h)^{m+q}}
\int_h^a\,...\,dy_i^{m+q} 
\end{eqnarray}
Later, additional averages over the particle numbers and the height h (which is equivalent to an average over the random shifts) will be performed.
The dashed line in fig. \ref{BOXPART} divides the two cells into four cell fractions; the
restricted position average, (\ref{AVERY}), takes into account that particles can be at any height $y$ within
their corresponding cell fractions with equal probability.
No averaging over the lateral positions is needed since the flow speed depends on $y$ but not on $x$.
For small collision parameter $A$ the acceptance probability is given by eq. (\ref{PACCEPT}) and is proportional
to $\theta(\Delta u)$. $\Delta u$ is equal to the difference of the mean velocities in the two subcells
projected onto the $\bsigma_1=\hat{x}$ direction,  
\begin{equation}
\label{DUDEF_COLL}
\Delta u=u_A-u_B={1\over N_A}\sum_{i=1}^{n+m} v_{i,x}-{1\over N_B}\sum_{i=n+m+1}^{n+m+p+q} v_{i,x}\,.
\end{equation}
The particle number in the left subcell is $N_A=n+m$, and $N_B=p+q$ is the particle number in the right subcell.
In order to simplify the integration over the $\theta$-function, the exponential representation from eqs. (\ref{GHCALC6},\ref{GHCALC7})
is utilized again.
$\Delta v_{i,x}$ is the change of x-component of a particle and follows from the collision rule (\ref{COLLHORI})
as
$\Delta v_{i,x}=2(u_x-v_{i,x})
$. Here, $u_x$ is the center of mass velocity of the two subcells together, $u_x=(v_{1,x}+...+v_{n+m+p+q,x})/(N_A+N_B)$.
Eq. (\ref{COLLHOR1}) becomes
\begin{eqnarray}
\nonumber
& &\langle{\Delta} p_x\rangle_{nmpq}^H= \\
\label{COLLHOR2}
& &2 A N_A N_B \left\langle \left\langle  
\int_0^{\infty}\,dc\int_{-\infty}^{\infty}{dk\over 2\pi} 
{\rm e}^{ik(\Delta u-c)}
\Delta u \,\sum_{j=1}^{n+p} (u_x-v_{j,x}) \right\rangle_V \right\rangle_Y
\end{eqnarray}
The velocity average is performed over the approximated non-equilibrium  
velocity distribution 
$f_{N,x}$,
defined in eq. (\ref{DEFFNX}) which still depends on the height $y_i$ of the particles.
This suggests the following variable transformation:
\begin{equation}
\label{V-TRAFO}
\hat{v}_{i,x}=v_{i,x}-\dot{\gamma}y_i\,.
\end{equation}
In addition, the auxiliary variable $c$ is changed into $\tilde{c}$ by the transformation
\begin{eqnarray}
\nonumber
\tilde{c}&=&c-\dot{\gamma}Q \\
\label{C-TRAFO}
Q&=&{y_1+...+y_{n+m} \over N_A}-{y_{n+m+1}+...+y_{n+m+p+q}\over N_B}
\end{eqnarray}
which moves the $y_i$-dependence from the exponent into the integral boundaries.
One obtains,
\begin{eqnarray}
\nonumber
& &\langle{\Delta} p_x\rangle_{nmpq}^H= \\
\nonumber
& &2 A N_A N_B\,\Gamma^{n+m+p+q}\, \left\langle  
\int_{-\dot{\gamma}Q} ^{\infty}\,d\tilde{c}\int_{-\infty}^{\infty}{dk\over 2\pi} 
\int_{-\infty}^{\infty}\, 
{\rm exp}[ik(\Delta \hat{u}-\tilde{c})] \right. \\
\nonumber        
& & {\rm exp}\left\{-{1\over 2 k_B T}(\hat{v}_{1,x}^2+\hat{v}_{2,x}^2+...+\hat{v}_{n+m+p+q,x}^2) \right\}
\,[\Delta \hat{u}+\dot{\gamma} Q]  \\
\label{COLLHOR2}
& & \left. \right[(n+p)\hat{u}_x-\sum_{j=1}^{n+p}\hat{v}_{j,x}+\dot{\gamma} R\left] \right\rangle_Y
\end{eqnarray}
with
\begin{equation}
\label{RDEF}
R={n+p\over N_A+N_B}\sum_{j=1}^{n+m+p+q}y_j -\sum_{j=1}^{n+p}\,y_j
\end{equation}
In order to decouple the velocity integrals, the argument of the exponential is transformed
into a sum of squares by
changing the velocity variables in the left subcell,
\begin{equation}
w_i=\hat{v}_{i,x}-{k_B T ik\over N_A}\;\;{\rm for}\;i=1...N_A 
\end{equation}
and the ones in the right subcell,
\begin{equation}
w_i=\hat{v}_{i,x}+{k_B T ik\over N_B}\;\;{\rm for}\;i=N_A+1...N_A+N_B\,.
\end{equation}
It is straightforward to perform the resulting Gaussian integrals over the new variables $w_i$, however, 
it is sufficient to only evaluate the term linear in the shear rate $\dot{\gamma}$.
There is also a shear independent term, but it will cancel out later in the final average over the dividing line.
This is to be expected since there should be no net momentum transfer in equilibrium where $\dot{\gamma}=0$.
The next higher order term must be cubic in the shear rate, since the viscosity 
(which is proportional to the momentum transfer divided by the shear rate) should not depend on the sign of the shear rate.
The relevant linear term has a simple structure, 
\begin{eqnarray}
\nonumber
& &\langle{\Delta} p_x\rangle_{nmpq}^H= O(1)+O(\dot{\gamma}^3)+\\
\label{AFTER_W_INTEG1}
& &2 A N_A N_B\,\dot{\gamma}\,k_B T\, \left\langle  
\int_{-\dot{\gamma}Q} ^{\infty}\,d\tilde{c}\int_{-\infty}^{\infty}{dk\over 2\pi} 
{\rm exp}\left(-ik\tilde{c}-{k^2\gamma k_B T\over 2}\right) ik\,\Omega\,,\right\rangle_Y \\
& &\Omega=Q\left({p\over N_A}-{n\over N_B} \right)+R\gamma
\end{eqnarray}
One can show that replacing the lower shear dependent boundary in the $\tilde{c}$-integral
by zero will only cause an error of order $\dot{\gamma}^3$ in the momentum transfer which is negligible.
With this modification the integrals over $k$ and $c$ have simple solutions, and one arrives at 
\begin{equation}
\label{AFTER_W_INTEG1}
\langle{\Delta} p_x\rangle_{nmpq}^H= 
A N_A N_B\,\dot{\gamma}\sqrt{2 k_B T\over \pi \gamma}\, \left\langle \Omega \right\rangle_Y 
+O(1)+O(\dot{\gamma}^3)
\end{equation}
The position average $\langle ... \rangle_Y$ defined in eq. (\ref{AVERY}) over particle heights give  
$\langle y_i \rangle_Y=h/2$ for particles {\it below} the dividing line with $i=1,...n$. 
{\it Above} the dividing line, where $i=N_A+1,...,N_A+p$, one finds $\langle y_i \rangle_Y=(a+h)/2$.
These results help finding the following averages,
\begin{eqnarray}
\nonumber
\langle Q \rangle_Y&=&{a\over 2}\left({m\over N_A}-{q\over N_B} \right) \\
\nonumber
\langle R \rangle_Y&= &{a\over 2} {(n+p)(m+q)\over N_A+N_B}\\
\label{OMEGA_AV}
\langle \Omega \rangle_Y&=&{a\over 2}
\left[ {2(pm+nq)+nm+pq\over N_A N_B} -{pq\over N_B^2} -{nm\over N_A^2} \right]
\end{eqnarray}
The numbers in the cell fractions, $n$ and $m$, as well as $p$ and $q$ are fluctuating even at fixed $N_A$ and $N_B$
and are approximately binomially distributed.
This leads to the following rules when performing the binomial average $\langle ... \rangle_b$ over the particle numbers, 
\begin{eqnarray}
\nonumber
\langle n \rangle_b&=&N_A\,w\,,\;\;\;\;\;\;\;\;\;\;\;\;\;\;\;\;\;\;\;\;\;\;\;\langle p \rangle_b=N_B\,w \\
\nonumber
\langle m \rangle_b&=&N_A\,(1-w)\,,\;\;\;\;\;\;\;\;\;\;\;\;\;\;\langle q \rangle_b=N_B\,(1-w) \\
\label{BINOM_AV}
\langle nm \rangle_b &=& (N_A-1)w(1-w)\,,\;\;\;\langle pq \rangle_b=(N_B-1)w(1-w)\,,
\end{eqnarray}
where $w=h/a$ is the relative height of the dividing line.
Using eqs. (\ref{AFTER_W_INTEG1}-\ref{BINOM_AV}) and
averaging over the height of the dividing line, $\langle ... \rangle_h=(1/a)\int_0^a ... dh$,
the average momentum transfer at fixed $N_A$ and $N_B$ is obtained, 
\begin{equation}
\label{AFTER_ALL_HORI}
\langle\langle{\Delta} p_x\rangle\rangle_{N_A,N_B}^H= 
{A\dot{\gamma}a\over 6}\sqrt{k_B T\over 2\pi \gamma}\,\left[ 4 N_A N_B+{(N_A-N_B)^2\over N_A N_B}\right]
+O(\dot{\gamma}^3)
\end{equation}

\begin{figure}
\begin{center}
\includegraphics[width=3.0in,angle=0]{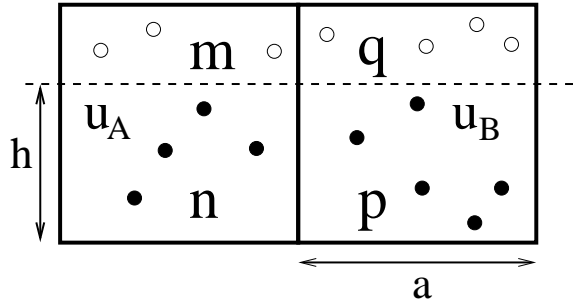}
\caption{
A double cell consisting of two subcells for horizonal collisions 
along the $\bsigma_1$ direction defined in fig. \ref{SUPERCOLL}.
$u_A$ is the horizontal component of the mean velocity of the $n+m=N_A$ particles in the left subcell.
The horizontal component of the $p+q=N_B$ particles in the right subcell is given by $u_B$.
}
\label{BOXPART}
\end{center}
\end{figure}

\subsection{Diagonal collisions}
Both diagonal collisions along $\bsigma_3$ and $\bsigma_4$ in fig. 1 will also transfer x-momentum over a horizonal diagonal line. Only the contribution from the upward diagonal, $\bsigma_3$, will be discussed here, since the
contribution from the down-wards diagonal gives the same result for symmetry reasons.
It is neccessary to distinguish between the two cases depicted in figs. \ref{DIAGBOXA} and \ref{DIAGBOXB}:
where (A) the dividing line goes through the lower cell and, (B) where the line goes through the upper cell.
Only both contributions together will give the correct result with the same structure than eq. (\ref{AFTER_ALL_HORI}).

First, case A is considered.
The change of x-momentum of the particles described by full circles in fig. \ref{DIAGBOXA} is given as 
\begin{equation}
\label{COLLDIAG1}
\langle{\Delta} p_x\rangle_{nmp}^A=\left\langle \left\langle  
p_A\,\sum_{j=1}^{n} \Delta v_{i,x} \right\rangle_V \right\rangle_Y
\end{equation}
where the upper index, $A$, denotes upward diagonal collisions for situation A.
The average over the particle height $\langle ... \rangle_Y$ and the velocity distribution $\langle ... \rangle_V$
differs from the previous definition eqs. (\ref{AVERV},\ref{AVERY})
\begin{eqnarray}
\label{AVERV1}
\langle ... \rangle_V&=&\int_{-\infty}^{\infty}\,...f_{N,x} f_{N,y}\,dv_{i,x}^{n+m+p}\,dv_{i,y}^{n+m+p} \\ 
\label{AVERY1}
\langle ... \rangle_Y&=&{1\over h^n}\int_0^h\,dy_i^n {1\over (a-h)^m}
\int_h^a\,dy_i^m 
{1\over a^p}\int_a^{2a}\,...\,dy_i^p
\end{eqnarray}
the velocity average now involves the $y$-component of the particle velocities as well.

The acceptance probability $p_A$ is again given by eq. (\ref{PACCEPT}) and proportional to
$\theta(\Delta u)$ which however has now a different form,
\begin{equation}
\label{DUDEF_COLL}
\Delta u=u_A-u_B={1\over N_A}\sum_{i=1}^{n+m} {v_{i,x}+v_{i,y}\over \sqrt{2}}-
{1\over N_B}\sum_{i=n+m+1}^{n+m+p} {v_{i,x}+v_{i,y}\over \sqrt{2}}\,.
\end{equation}
The particle number in the bottom left cell is $N_A=n+m$, and $N_B=p$, is the particle number in the top right cell.
Using the exponential representation (\ref{GHCALC6},\ref{GHCALC7}) for the $\theta$-function,
and the upward collision rule (\ref{COLLUDIAG}) to redefine $\Delta v_{i,x}=u_x+u_y-v_{i,x}-v_{i,y}$, one finds
\begin{eqnarray}
\nonumber
& &\langle{\Delta} p_x\rangle_{nmpq}^A= A N_A N_B \\
\label{COLLDIAG2}
& &\times \left\langle \left\langle  
\int_0^{\infty}dc\int_{-\infty}^{\infty}{dk\over 2\pi} 
{\rm e}^{ik(\Delta u-c)}
\Delta u \,\sum_{j=1}^{n+p} (u_x+u_y-v_{j,x}-v_{j,y}) \right\rangle_V \right\rangle_Y
\end{eqnarray}
This expression is evaluated using similar substitutions and approximations than for horizontal collisions.
After performing all averages one obtains the momentum transfer for fixed particle numbers $N_A$, $N_B$
(in linear order in the shear rate $\dot{\gamma}$):
\begin{equation}
\label{AFTER_ALL_A}
\langle\langle{\Delta} p_x\rangle\rangle_{N_A,N_B}^A= 
{A\dot{\gamma}a\over 12}\sqrt{k_B T\over 2\pi \gamma}\,
\left[(N_A-1)\left(1-{N_B\over N_A}\right)+14 N_A N_B  \right]
\end{equation}
A similar calculation for case B where the dividing line cuts the top right cell, see fig. \ref{DIAGBOXB}, results in
\begin{equation}
\label{AFTER_ALL_B}
\langle\langle{\Delta} p_x\rangle\rangle_{N_A,N_B}^B= 
{A\dot{\gamma}a\over 12}\sqrt{k_B T\over 2\pi \gamma}\,
\left[(N_B-1)\left(1-{N_A\over N_B}\right)+14 N_A N_B  \right]
\end{equation}
By averaging both results, one finds the momentum transfer for the upward (U) diagonal collision
for $N_A\ge 1$ and $N_B\ge 1$:
\begin{eqnarray}
\nonumber
\langle\langle{\Delta} p_x\rangle\rangle_{N_A,N_B}^U&=&{1\over 2} 
\left( \langle\langle{\Delta} p_x\rangle\rangle_{N_A,N_B}^A 
+\langle\langle{\Delta} p_x\rangle\rangle_{N_A,N_B}^B 
\right) \\
\label{ALL_UPWARDS}
&=&
{A\dot{\gamma}a\over 6}\sqrt{k_B T\over 2\pi \gamma}\,\left[ 7 N_A N_B+{(N_A-N_B)^2\over 4 N_A N_B}\right]
\end{eqnarray}
This expression is symmetric in $N_A$ and $N_B$, as it should be, and has the same general structure than eq. (\ref{AFTER_ALL_HORI}).
Finally, this result is multiplied by two to account for the contribution from the downward diagonal
and added together with the result for horizontal collisions using the weight factors $w$ and $w_d$ for
horizontal and diagonal collisions, respectively.
The total average momentum transfer is:
\begin{eqnarray}
\nonumber        
& &\langle\langle{\Delta} p_x\rangle\rangle_{N_A,N_B}= \\ 
\label{ALL_TOTAL}
& &{A\dot{\gamma}a\over 12}\sqrt{k_B T\over 2\pi \gamma}\,\left[ (28 w_d+8w) N_A N_B+(w_d+2w){(N_A-N_B)^2\over N_A N_B}\right]
\end{eqnarray}
Now, one can derive the collisional part to the kinematic viscosity, $\nu_{coll}$, which is proportional to the momentum transfered
per time and length
\begin{equation}
\label{VISCO_DEFINI}
{ \langle \Delta p_x\rangle \over 2a\,\tau}=\dot{\gamma}\, \nu_{coll} \,\rho 
\end{equation}
$\rho=M/a^2$ is the density of our two-dimensional fluid.
Replacing the particle numbers $N_A$ and $N_B$ by the average particle number per cell, $M$ and setting
$w=1/4$ and $w_d=1/2$,
one gets
\begin{equation}
\label{VISCO_COLL}
\nu_{coll}={a^2\over 3\tau} A\sqrt{k_B T\over \pi}\,M^{3/2}
\end{equation}
for not too small mean particle number per cell, $M\ge 2$.
Figs. \ref{VISCOS_A1-60_1} and \ref{VISCOS_A1-60_2} compare the theoretical result for the total viscosity, $\nu=\nu_{coll}+\nu_{kin}$
from eqs. (\ref{VISCOS_KIN_A},\ref{VISCO_COLL}) with numerical results obtained by evaluating Green-Kubo relations
for stress autocorrelation functions (for numerical details see refs. \cite{ihle_06,ihle_08}).
One sees that the agreement at small and large mean free paths is very good.
The small deviations at intermediate mean free path are within the error bars of the data.
However, there is another possibility; 
in regular SRD it was found, \cite{tuze_06c}, that at $\lambda\ll a$ corrections due to the violation of molecular chaos become relevant
in $\nu_{kin}$, but not in $\nu_{coll}$. These effects are large for very small $\lambda$, 
but in this limit they are irelevant for the total viscosity because it is dominated by $\nu_{coll}$.
Hence, if at all, one would see these deviations only at intermediate mean free paths.
It is possible that there are similar deviations for the current model which could show at intermediate $\lambda$.

There is a simple argument for the structure of expression (\ref{VISCO_COLL}):
Collisions ``smear out'' momentum over a distance of order cell size $a$ in one time step $\tau$.
The kinematic viscosity is the coefficient of momentum diffusion
which in analogy to a random walk is defined by the ``hopping '' distance $a$ per time step as
$D\sim a^2/\tau$.
However, the collisions occur with some collision rate which is proportional to
the thermal average of the acceptance probability given by
\begin{equation}
\label{PA_AVER}
\langle p_A\rangle=A\,\langle \theta(\Delta u)\,\Delta u N_A N_B\rangle
\sim A\,\sqrt{k_B T}\, M^{3/2}
\end{equation}
Therefore, one would expect
\begin{equation}
\label{DEGENNES_COLL}
\nu_{coll}\sim {a^2\over \tau}\langle p_A\rangle \sim {a^2\over \tau}\,A\,\sqrt{k_B T}\, M^{3/2}
\end{equation}
which is exactly what was found.
A similar argument can be made for the kinetic part of the viscosity.
The analogy to a random walk predicts $\nu_{kin}\sim \lambda^2/\tau$
since momentum is now only ``hopping'' a distance of order mean free path $\lambda=\tau\sqrt{k_B T}$.
In addition, it is clear
that at $p_A\rightarrow 0$ there is almost no 
collisions, the momentum is travelling huge distances and the kinetic viscosity becomes very large.
This can be described by an additional factor $1/\langle p_A \rangle$ and one obtains,
\begin{equation}
\label{DEGENNES_KIN}
\nu_{kin}\sim {1\over \langle p_A \rangle} {\lambda^2 \over \tau}\sim {k_B T \tau\over \langle p_A \rangle}
\sim {\tau\over A} \sqrt{k_B T} \, M^{-3/2}
\end{equation}
which, apart from a small correction term, agrees with what was derived rigorously, eq. (\ref{VISCOS_KIN_A}).
Using this conceptual insight, one can easily predict
the general form of these expressions for arbitrary acceptance probabilities without actually going through
tedious derivations.

\section{Results for other models}

\subsection{Large collision rate, $A\rightarrow \infty$}

The calculation of the previous sections are easily modified to treat the limit of large collision parameter
$A\rightarrow \infty$, where the acceptance probability is simply $p_A=\theta(\Delta u)$.
One finds,
\begin{equation}
\label{VISCO_AI_KIN}
\nu_{kin}^{\infty}= {k_B T \,\tau\over 2}\,\left({6M+1-{\rm e}^{-2M}\over 2M-1+{\rm e}^{-2M}}\right) 
\end{equation}
for  the kinetic viscosity if ideal gas-like fluctuations are assumed.  
The exponential terms come from averaging over the particle number fluctuations assuming a Poisson-distribution.
This assumption is correct for an ideal gas, but not for the current model with a non-ideal equation of state, which has smaller fluctuations.
Furthermore, it was shown that the density fluctuations in the limit of large $A$ are even smaller than predicted by the equation of state \cite{tuze_06b}.
Given the small size of the exponential terms $\sim {\rm e}^{-2M}$ for typical particle numbers $M$ between three and twenty, 
the fluctuation corrections, i.e. the exponential terms, can be safely neglected.

By modifying the non-equilibrium calculation scheme of section 5, one obtains the average momenum transfer in shear flow for $N_A\ge 1$, $N_B\ge 1$
as
\begin{eqnarray}
\nonumber        
& &\langle\langle{\Delta} p_x\rangle\rangle_{N_A,N_B}= \\ 
\label{ALL_TOTAL_AI}
& &{a \dot{\gamma}\over 24(N_A+N_B)}\,
\left[ (14 w_d+4w)N_A N_B+(w_d+2w)(N_A+N_B-2) \right]
\end{eqnarray}
Ignoring particle number fluctuations by setting $N_A=N_B=M$, substituting $w=1/4$, $w_d=1/2$, 
the collisional viscosity 
follows from eq. (\ref{VISCO_DEFINI}),  
\begin{equation}
\label{VISCO_AI_COLL}
\nu_{coll}^{\infty}={a^2\over 12\tau}\left[1+{1\over 4M}\left(1-{1\over M}\right) \right] 
\end{equation}
Fig. \ref{VISCOS_AINFIN} shows excellent agreement of simulation data for $M=3$ from ref. 
\cite{tuze_06b} with the theoretical results for the total viscosity $\nu=\nu_{kin}+\nu_{coll}$ from eqs. 
(\ref{VISCO_AI_KIN}, \ref{VISCO_AI_COLL}).
One sees that the viscosity is basically independent of the particle number $M$ at small mean free path.

\subsection{Transport coefficients for a binary mixture}

In \cite{tuze_07} a multi-component version of the current collision model was introduced.
Consider a binary system with  two types of particles, A and B.
In order to obtain phase separation, a repulsion between different kind of the particles was implemented; but there is no 
repulsion among particles of the same kind.
This is done in the following way: Suppose a double cell is selected for a possible collision. A-particles in cell 1 can undergo collisions with B-particles from cell 2. For symmetry reasons, 
B-particles from cell 1 are also checked for possible collision with A-particles in cell 2.
The rules and probabilities for these collisions are exactly the same as in the one-component situation.
This means, most of the results derived above can be used with only minor modifications. 

$M_A=\langle N_A \rangle$ is the average number of A-particles in a cell of size $a$, 
$M_B=\langle N_B \rangle$ is the average of B-particles in such a cell.
The total particle density in a cell is now given by $\rho=(M_A+M_B)/a^2$.
While using eqs. (\ref{GHCALC2}) to (\ref{GHCALC12}) for diffusion calculations one has to distinguish whether the tagged particle is of the 
A or B type. 
For the self-diffusion of A-particles one has
to set $n=N_A$ and $m=N_B$ in eq. (\ref{GHCALC4}) for $K^l_{nm}$,
but $n=N_B$ and $m=N_A$ in $K^r_{nm}$ which, because of the symmetry $K^r_{nm}=K^l_{mn}$, leads to
$K_{nm}=K^l_{N_A N_B}$.
The corresponding results for the auxiliary variables $g_H$, $g_U$ and $g$ are not symmetric with respect
to the particle numbers $N_A$ and $N_B$, and a short calculation gives 
$g_H^A=1-2\phi^A$,  
$g_U^A=1-\phi^A$, and
$g^A=1-\phi^A$ with 
\begin{equation}
\label{G_FOR_BIN}
\phi^A=A\sqrt{k_B T\over 2\pi \gamma}\left(M_A+M_B-1+{M_B\over M_A}\right)
\end{equation}
and $\gamma=1/M_A+1/M_B$ where the actual numbers $N_A$, $N_B$ were already replaced by their average values, 
$M_A$ and $M_B$, respectively.
Similar considerations hold for the diffusion of B particles and one obtains the diffusion coefficient
for A and B particles, respectively,
\begin{eqnarray}
\nonumber         
D_A&=&k_B T \tau\left({1\over \phi_A}-{1\over 2} \right) \\
\nonumber         
D_B&=&k_B T \tau\left({1\over \phi_B}-{1\over 2} \right)\;\;{\rm with} \\ 
\label{DIFFUS_BINARY}
\phi^B&=&A\sqrt{k_B T\over 2\pi \gamma}\left(M_A+M_B-1+{M_A\over M_B}\right)
\end{eqnarray}
Both particles diffuse the same way only if $M_A=M_B$. In this case the diffusion coefficient $D=D_A=D_B$
agrees with the one for the single-component model, eq. (\ref{EXACT_DIFF}).
If there is less A than B particles, the diffusion of A particles is slower even if the masses of both particles
are the same as assumed here.
This is physically plausible, since a single A-particle in a sea of B particles will be backscattered 
very often whereas the many B-particles keep going and are less affected by the collisions.
Fig. \ref{DIFFUS_RATIO_BIN} plots the ratio of the two diffusion coefficients $D_A/D_B$ as a function
of the averaged relative particle number difference, 
$\Delta \rho=(M_A-M_B)/( M_A+M_B)$ for fixed total number $M=M_A+M_B$.
$\Delta \rho$ was only varied in a range were both particle numbers $M_A$ and $M_B$ are always larger or equal to one
because density fluctuations become significant at smaller particle numbers but are neglected in the theory. 
For a total number of $M=5$ and $M_A=1$, $M_B=4$ one sees
that B-particles diffuse about twice as fast than A-particles.

Finally, the kinetic and collisional viscosities are determined by applying concepts from the previous sections,
and one finds
\begin{eqnarray}
\nonumber
\nu_{kin}^{bin}&=&k_B T\,\tau\left\{{1\over A}\sqrt{\pi\over 2 k_B T} [M_A M_B(M_A+M_B)]^{-1/2} -{1\over 2}\right\} \\
\label{VISCOSITY_BIN}
\nu_{coll}^{bin}&=&{Aa^2\over 3\tau}\sqrt{k_B T M_A M_B\over 2\pi (M_A+M_B)^3}
\left[ 4 M_A M_B +{(M_A-M_B)^2\over 4 M_A M_B}\right]
\end{eqnarray}

\begin{figure}
\begin{center}
\includegraphics[width=3.0in,angle=0]{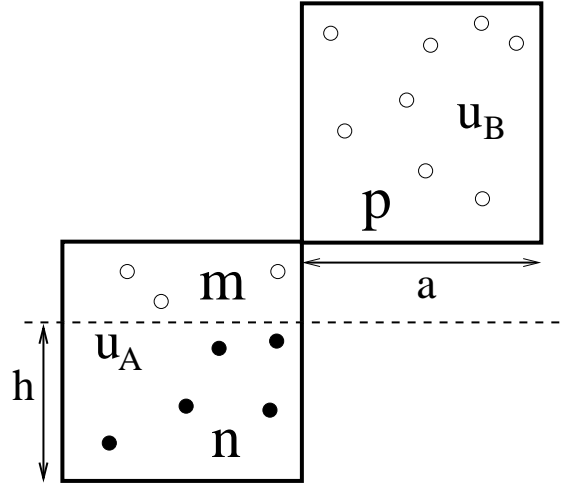}
\caption{
A double cell consisting of two subcells for collision 
along the upward diagonal direction $\bsigma_3$ as defined in fig. \ref{SUPERCOLL}.
$u_A$ is the component of the mean velocity projected on $\bsigma_3$
 of the $n+m=N_A$ particles in the left subcell.
The projected mean velocity of the $p=N_B$ particles in the upper right subcell is given by $u_B$.
A line divides the lower cell at height $h$, $0\le h \le a$.
}
\label{DIAGBOXA}
\end{center}
\end{figure}
\begin{figure}
\begin{center}
\includegraphics[width=3.0in,angle=0]{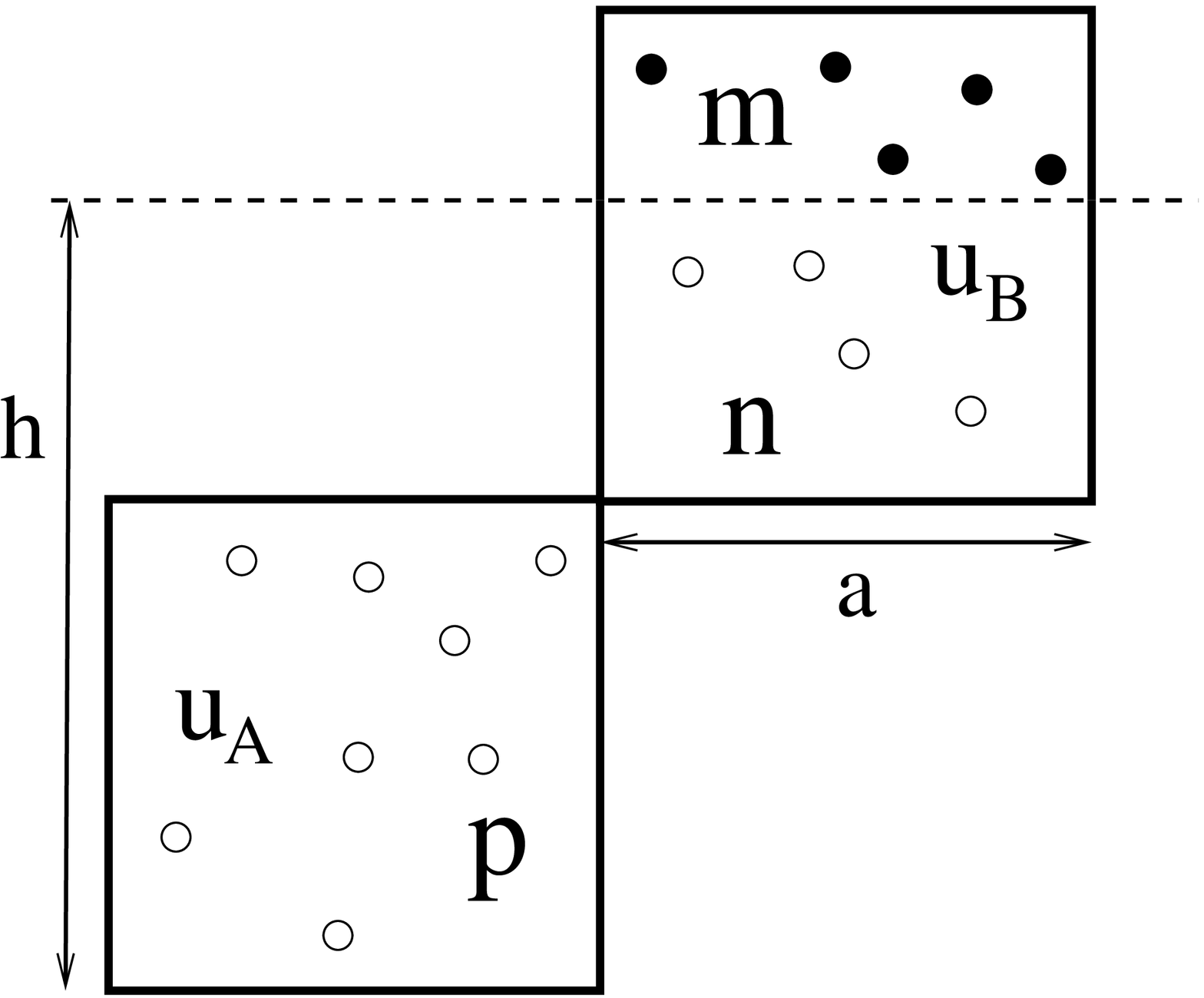}
\caption{
A double cell consisting of two subcells for collision 
along the upward diagonal direction $\bsigma_3$ as defined in fig. \ref{SUPERCOLL}.
$u_A$ is the component of the mean velocity projected on $\bsigma_3$
 of the $p=N_A$ particles in the lower left subcell.
The projected mean velocity of the $n+m=N_B$ particles in the upper right subcell is given by $u_B$.
A line divides the upper cell at height $h$, $a\le h \le 2a$.
}
\label{DIAGBOXB}
\end{center}
\end{figure}

\begin{figure}
\vspace{0.5cm}
\begin{center}
\includegraphics[width=3.0in,angle=0]{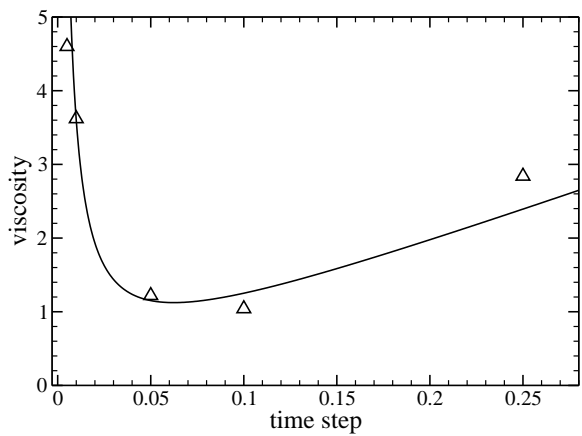}
\caption{
Total viscosity $\nu$ as a function of time step $\tau$ for small collision rate.
The symbols are numerical results obtained by evaluating Green-Kubo relations for $M=3$.
The solid line is the theoretical result, $\nu=\nu_{coll}+\nu_{kin}$ from eqs. 
(\ref{VISCOS_KIN_A},\ref{VISCO_COLL}).
Parameters: $L=64a$, $M=5$, $k_B T=1$, and $A=1/60$.
}
\label{VISCOS_A1-60_1}
\end{center}
\end{figure}
\begin{figure}
\vspace{0.5cm}
\begin{center}
\includegraphics[width=3.0in,angle=0]{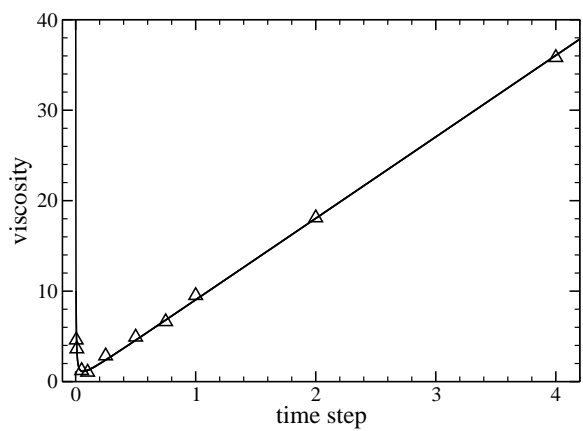}
\caption{
Same as fig. \ref{VISCOS_A1-60_1}: total viscosity $\nu=\nu_{coll}+\nu_{kin}$ as a function of time step $\tau$, but for larger $\tau$, and compared to the theoretical result. 
}
\label{VISCOS_A1-60_2}
\end{center}
\end{figure}

\begin{figure}
\vspace{0.5cm}
\begin{center}
\includegraphics[width=3.0in,angle=0]{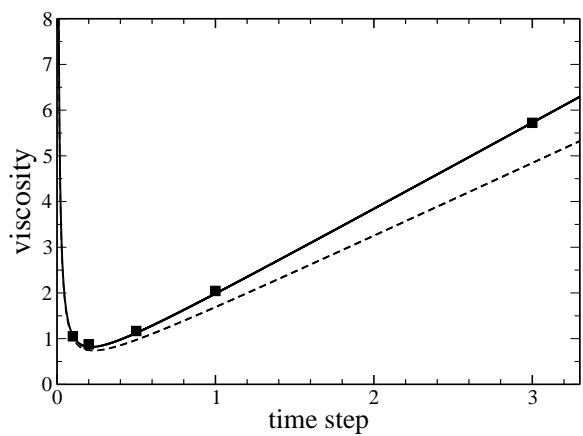}
\caption{
Total viscosity $\nu=\nu_{coll}+\nu_{kin}$ as a function of time step $\tau$ for large collision rate, $A\rightarrow \infty$.
The symbols are numerical results obtained by evaluating Green-Kubo relations for $M=3$.
The solid line is the theoretical result from eqs. (\ref{VISCO_AI_KIN},\ref{VISCO_AI_COLL}), for $M=3$.
The dashed line is the theoretical result for $M=10$.
Parameters: $L=64a$, $k_B T=1$, and $A\rightarrow \infty$.
}
\label{VISCOS_AINFIN}
\end{center}
\end{figure}

\begin{figure}
\vspace{0.5cm}
\begin{center}
\includegraphics[width=3.0in,angle=0]{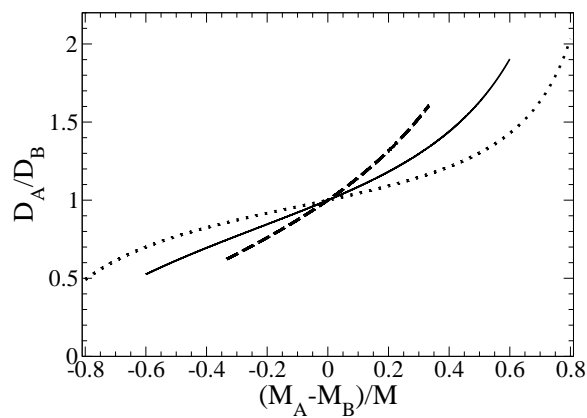}
\caption{
Ratio of the diffusion coefficient for A and B particles, $D_A/D_B$, as a function of averaged relative
particle number $\Delta \rho=(M_A-M_B)/M$ for fixed total number $M=M_A+M_B$.
The dashed line is the theoretical result from eqs. (\ref{DIFFUS_BINARY}), for $M=3$.
The solid line is for $M=5$, and the dotted one is for $M=10$.
Parameters: $k_B T=1$, $A=1/60$, $\tau=1$, $M_A\ge 1$, $M_B\ge 1$.
}
\label{DIFFUS_RATIO_BIN}
\end{center}
\end{figure}

\section{Conclusion}
In this paper, I have presented a detailed, systematic derivation of the transport coefficients
of a multi-particle collision model for fluid flow with a non-ideal equation of state.
Similar calculations have been published before, but to my knowledge, this is the first derivation which accounts for
velocity-biased collision rules in MPC.
Analytic expressions for the self-diffusion coefficient, and the shear viscosity are obtained, 
and very good
agreement is found with
numerical results at small and large mean free paths.
For the analysis of the collisional contribution to the viscosity, an earlier scheme was improved which can now be 
used to derive possible shear thinning or thickening.
The general formalism was also applied to a binary collision model with a miscibility gap.
General arguments for the scaling of the transport coefficients with temperature and particle number 
as a function of the collision rule are given.
This allows one to predict without tedious derivations
how the transport coefficients change if the collision rule is modified.

For the current collision rule, the viscosity turns out to be  proportional to the square root of temperature as in a real gas.
It is shown that at large mean free path, the viscosity and the diffusion coefficient are exactly equal,
resulting in a Schmidt number of one in this limit.

\section{Acknowledgement}
Acknowledgement is 
made to the Donors of the American Chemical Society Petroleum Research Fund for partial support of
this research.
I thank F. J{\"u}licher for his hospitality at 
the MPI-PKS. 
I would like to thank E. T{\"u}zel for discussions and help in the initial stage of this work, and  
D. Kroll and T. Pilling for many discussions and proof-reading of the manuscript.
I also thank J. Yeomans for providing unpublished results about shear-thinning.

\end{document}